\documentclass[twocolumn,showpacs,preprintnumbers,amsmath,amssymb,epsf,prl,pdftex,a4paper]{revtex4}

\usepackage{graphicx}
\usepackage{dcolumn}
\usepackage{bm}
\usepackage{color}
\usepackage{portland}
\usepackage{ulem}
\usepackage{float}
\newcommand{\be}{\begin{eqnarray}}
\newcommand{\ee}{\end{eqnarray}}

\newcommand{\bs}{\bm}

\usepackage{setspace}

\begin{document}

\title{B-H hysteresis in itinerant Ferromagnetism  from Chern-Simons Gauge theory }

\author{Kenzo Ishikawa}
\email{ishikawa@particle.sci.hokudai.ac.jp}
\affiliation{Department of Physics, Faculty of Science, Hokkaido University, Sapporo, 010-0810, Japan}
\affiliation{Natural Science Center, Keio University, Yokohama 223-8521, Japan}

\date{\today}


\begin{abstract}

Applying  the Chern-Simons (CS) gauge theory for itinerant ferromagnetism,  we  demonstrates that  the universal $\log H$ term in the free energy leads the electron system to the first order transition.  Singular behavior  at $H=0$ induces  the separation of  positive and negative magnetzations  and the $B$-$H$ hysteresis.  We thus establish  that intriguying quantum phenomena of  itinerant ferromagnetism  of the single and symmetric domain crystall.  This mechanism naturally accounts for a broad class of itinerant ferromagnets, revealing the gauge-theoretic origin of magnetism. 

\end{abstract}

\maketitle
{\it Introduction}.Itinerant ferromagnets have long been of interest for both their role in correlated electron physics and their potential applications in wide modern  technologies~\cite{Wolf, Zutic}. Yet the microscopic mechanism of itinerant ferromagnetism remains unclear.  The Stoner model~\cite{Stoner} captures  a basic mean-field picture, and the Hubbard model~\cite{Hubbard} stabilizes ferromagnetism  in restricted parameter regimes, and spin-fluctuation theories~\cite{Moriya} account for some anomalies in weak itinerant ferromagnets such as ZrZn$_2$ and MnSi. First-principle approaches~\cite{Georges, Held} reproduce experimental trends only partially.

The $B$-$H$ hysteresis   in a single and symmetric  domain is one of  challenging topics for theory and  applications.   Non-invertible deformation which can produce hysteresis in multiple  domains is absent in the single domain. Rotational motion in the asymmetric interaction energy of the magnetization with the crystall structure was incorpolated in  Stoner-Wohlfarth model \cite{Stoner_2}.   The present paper elusidates that the inversion of the magnetization  reveals  the quantum phase transition of the  first order.         

In a recent publication,  $\log H$   term was found  in the free energy  \cite{Ishikawa-NPB}. This has an origin in the intrinsic nature of the electron wave function in the magnetic field and represents a long distance fluctuation, which is  expressed by  the Chern-Simons  term of the coupling strength invesly proportional to the magnetic field. Due to  unique properties, this can spontaneously break time-reversal symmetry and induces finite magnetization.  We show that this framework consistently accounts for $B$-$H$ hysteresis in a single domain.  Throughout, we adopt natural units with 
$ \hbar = c = \epsilon_0=\mu_0=1$~\cite{Footnote-charge}, and $\bs H$ and $\bs B$ for the external magnetic field and induced flux. 

{ \it  Free energy $f(H)$, quantum phase transition,  and spontaneous magnetization}.
We express the   applied field $ \bs H$ with the external vector potential by $\bs H= \nabla \times \bs A_{ext}$ and compute  the energy density of the many-electron system  $f(\bs H)$. Details of the calculation will be presented in a separate publication.    The energy density  proportional to ${\bs H}^2$,  $f_0$, is  
\begin{eqnarray}
 f_0=  \frac{1+ \chi}{2}  {\bs H}^2, \label{field-energy}
 \end{eqnarray}
where  $\chi$ is zero in the vacuum.  $\chi$ depends on material and is assumed constant.   Chern-Simons  term $A_0 B_i$ emerging from  the long-range correlation  was computed in  
Ref.\cite{Ishikawa-NPB}  using the effective Lagrangian method for the isolate state in which  the vector fields vanish at the spatial infinity.  The long-range component is sensitive to the boundary condition in  symmetric systems.  If the filed does not vanish at boundary, spurious Goldstone modes contribute from non-orthogonality ~\cite{Orthogonality,Mahan}. To compute the  energy of isolate system, it is necessary to separate  spurious Goldstone mode.  In the Landau gauge $(A_x,A_y)=B(-y,0)$,this problem arizes in the $x$ direction and a spurious would-be Goldstone mode exhibits pathological 
behavior, resulting in an unphysical outcome \cite{Sin-Itiro-Tomonaga,Thouless-Valatin}.  
This problem can be avoided by adding a regularizing term  $\frac{\epsilon x^2}{2}$ into the potential. First compute the eigenstates with a finite $\epsilon$ and compute the total energy. By taking  the limit $\epsilon \rightarrow 0$ at the end, the regulaized value in which the spurious states decouple is computed. This agrees with ~\cite{Footnote-symmetric-gauge, von-Neumann}. See also \cite{Footnote-variation-free-energy, Footnote-TF} .
The differential of free energy $\delta f$  is  consistent with   
  Ref.~\cite{Ishikawa-NPB}. 




 We obtain    the  free energy  by integrating
  $\delta f$,  over the gauge-flux insertion 
$\delta A_\mu$ in the limit of $T\rightarrow 0$, and study physical implications.
Omitting the $H$-independent contributions, the result is 
\be
f_1(H)&=&-C_0 \int_H \frac{\delta H'}{H'}
=-\ln \frac{ |H|}{H_0} C_0 
\label{free_B}
\ee
where  $e \delta A_\tau=i \delta \mu$ ,$ C_0 ={\int_\mu n_e c_\Gamma \delta \mu' }$. Here   $\Gamma$ is the width and   $c_\Gamma$ is the correction term that depends on the width, and $c_{\Gamma=0}=1$.   Integration constant $H_0$  will be discussed later.   
In the limit $\Gamma \rightarrow 0$, the result reproduces that of  Ref. \cite{Ishikawa-NPB}  up to a numerical factor. 

The spin degree of freedom does not affect  Eq.$(\ref{free_B})$, but modifies  the susceptivity of 
 Eq.~(\ref{field-energy})  as
$ f_s= \frac{ \chi_s}{2}  {\bs H}^2 $
where $\chi_s$ is the spin susceptibility. For simplicity, we focus  the case  no  H-linear term in the following at low temperature ~\cite{Footnote-B-linear}.
Adding $f_0$, $f_1$, and $f_s$, and using $\chi=1+\chi_s$, we have  free energy 
\begin{eqnarray}
 f(H)=   -C_0 \log \frac{|H|}{H_0}+\frac{ 1+ \chi}{2}  {\bs H}^2. \label{free-energy}
 \end{eqnarray}
 
 For simplicity, we focus on the system at low temperature, where $\chi$ is the constant. The first term in  $f(H)$ is divergent  at $H=0$, and the motions at   $H>0$ and  $H<0 $ are separated in classical systems. Quantum tunneling makes two processes connected and governs physical constants  in two branches, $H_0(\pm)$. This is equivalent to the  renormalization \cite{Schwinger}. These   depend upon the integration path  owing to the singularity at $H=0$. We evaluate these  from  the zero-point energy   of the state of adiabatical proess of changing    $H$   from a large positive value to a large negative value.  In the first half of climbing the energy hill, the quantum tunneling takes place and makes a finite  energy to move  from  $H>0$ to  $H<0$. The tunneling  probability   and average energy  $\Delta >0$  are evaluated  in Appendix.  In the second half of going down the hill, the tunneling does not take place.  Because energy $\Delta$ is moved from $H>0 $ to $H<0$, two energies are connected by $f_{-}(|H|) = f_{+}(|H|) +2\Delta$.

   To describe the adiabtic change crossing $H=0$, as shown in Appendix, we can use  the conjugate variable,
induced magnetic flux    given by  
\be
B=\frac{\partial f}{\partial H}=-\frac{1}{H} C_0  +(1+\kappa)H.
\label{B-H}
\ee
The field strength in each branch is expressed  by 
\be
 H_{\pm}=\frac{1}{2(1+\kappa)}[B  \pm \sqrt{B^2+ 4 C_0 (1+\kappa)}],
\label{H-B}
\ee
and   Gibbs free energy $g(B)=BH-f(H)$ is expressed by
\begin{eqnarray}
& &g_{\pm}(B)=\frac{1}{4(1+\kappa)}[B^2  \pm B \sqrt{B^2+ 4 C_0 (1+\kappa)}]-\frac{C_0}{2}  \nonumber \\
& &+C_0 \log  [\frac{1}{2(1+\kappa)H_0(\pm)}|B  \pm \sqrt{B^2+ 4 C_0 (1+\kappa)}|]. \label{gfree-energy}
\end{eqnarray}
$g_{\pm}(B)$ does not diverge at finite $B$. Conversly,   $H=\frac{g(B)}{\partial B}$. 

 Two branches   intersect at  $B_c$ of satisfying  $g_{+}(B_c)=g_{-}(B_c)$, where for small tunneling energy, $\frac{\Delta}{f(H)} \ll 1$, $B_c=\sqrt{C_0(1+\kappa)}\Delta$.  Because, $g_{+}'(B_c) \neq g_{-}'(B_c)$, this is  a first order phase transition. 
 The intersection  region is  a line in $B-H$ space,  $B=B_c, H_{-}(B_c) \leq H \leq H_{+}(B_c)  $, 
\begin{eqnarray}
H_{\pm}(B_c)=\pm \sqrt{\frac{C_0}{1+\kappa}}( \sqrt{1+\frac{\Delta^2}{4}}+\frac{\Delta}{2})
\end{eqnarray}
which includes $H=0$. $B_c$ is the induced flux at $H=0$, i.e.,  spotaneous magnetization.  $H_c(0)=- \sqrt{ \frac{C_0}{1+\kappa}} $ is the  coercive force \cite{Footnote-coercive}.  Area of the enclosed region   $ \sim (H_{+}(B_c)-H_{-}(B_c))B_c$ is the supplied energy during the  motion.    
The adiabatic motion at $H >H_{+}(B_c)$ and $H< H_{-}(B_c)$ can be described by $f(H)$ also.   While $H$ is decreased adiabatically from the $H>0$ region, that follows  to $f_i(H)$ up to critical strength $H_c=H_{+}(B_c)$.
   
   Transions of $H$   in the   region  $H_{-}(B_c) <H< H_{+}(B_c)$  governed by  quantum mechanics  are consistent with an increase  of the uncertainty of the position and momentum  in the wave-packet scatterings  \cite{Ishikawa-Tobita,Ishikawa-Nishio}

{\it B-H curve and Hysteresis of single crystall  }
 
 When $H$ is increased  adiabatically from  $H <0$  to $H>0$, the system' motion is discribed by  $g_{-}(B)$ till that intersects with  $g_{+}(B)$ of  $f_{+}(|H|)=f_{-}(|H|)+2 \Delta$.  After the intersection, that follows $g_{+}(B)$.  The latter slope  is different from the previous slope, and this corresponds to the first order transition.  Whole trajectory is shown in Figure, and the area of closed region  is finite for $\Delta \neq 0$ and approximately expressed as $2 \times (H_{+}(B_c)-H_{-}(B_c))B_c $ for $\Delta \sim 0$.  The field supplies this amount of energy to outsize during adiabatic process.  The spontaneous magnetization and hysterisis are caused by the tunneling energy $ \Delta >0 $.  
\\

The  present mechanism of the first order transition and the $B$-$H$  hysteresis  in the single domain is caused by the quantum mechanical dynamics. This is irrelevant to the coherent rotation of the magnets   and to the  non-invertible motions of  domains.   The mechanism works in the uniform systems of the single domain. Single domain magnets including planer magnets are advancing rapidly, and it would be exciting to verify the present mechanism experiemtally.

The present theory is readily seen to yield a first-order transition (as will also be the case in 3D systems, as discussed later) independent of the temerature, which  is not related with the divergence of  entropy as $T \rightarrow T_c^{-}$.

 Although a first-order transition also appears in conventional fluctuation-based theories of itinerant ferromagnetism~\cite{Moriya,Belitz-PRL99}, many experiments  show that the associated singularity is smeared out in realistic systems by sample-dependent factors, so that only an apparently second-order behavior is observed~\cite{Brando-RMP,Goko-npjQM}.   
Hysteresis at the crossing $ H =0$  in our theory,  on the other hand,  appears  always in a single domain.
The logarithmic divergence of the free energy at $H=0$
causes  hysteresis in the bare theory. This is caused by the specific property of  wave functions in the magnetic field, and  is topologically stable. This is neither  smeared  out  by  impurities  nor  cancelled by higher order corrections.
Therefore, this divergence may not be   cut off by sample-dependent mechanisms (e.g., finite-size effects or inhomogeneity). Nevertheless, $H_{\pm}(B_c)$ and $H_c(0)$ become huge in the system $\kappa \sim 0$. Accordingly this mechanism is  rellevant to itinerant magnetism  and  various parameters in many domains may receive corrections consistent with the extrinsic and strongly sample-specific nature of the coercive force~\cite{Kittel}. 
Recent progresses in single domain experiments show that the hysteresis reveals universal features which are consistent with the results of the present paper. 




{\it Extension to 3D systems}
The above discussions would apply to 2D itinerant ferromagnets~\cite{Fe3GeTe2}, and 
also operate in three-dimensional (3D) systems. 
The CS term in 3D, or Carol--Field--Jackiw term 
\be
\mathcal{L}_{CFJ}=
\frac{\kappa_{CFJ}}{2}\epsilon^{\mu\nu\rho\sigma}n_\mu A_\nu \partial_\rho A_\sigma 
\ee
with a unit vector $n_{\mu=0,x,y,z}=(0, \vec{B}/|\vec{B}|$) \cite{Carol-Field-Jackiw}, is induced from the 1-loop contribution of the 3D fermion, 
and its coefficient $\kappa_{CFJ}=e n_e/ |\vec{B}|$, i.e., the singularity exits even in 3D 
(see, Eq. (121) in Ref.~\cite{Ishikawa-NPB}). The term, therefore, gives rise to 
the local expectation value $\vec{B}=\vec{B}_c$. 
In 3D the ferromagnetic condition loosens markedly, as the condensation energy outweigh 
surface energies (domain-wall, and demagnetization at sample surface~\cite{Kittel}).

{\it Summary and Discussions}.
In summary, we have proposed a universal mechanism for itinerant ferromagnetism that originates solely from the CS term, independent of microscopic band structures and spin exchange interactions. A key effect of the present mechanism is caused by   the  free energy  that is  proportional to $ \log H$. This  is  natural  consequence  of the many-electrons in the Landau levels where  the spatial sizes of  electrons are  inversly proportional to the magnetic field. The  unbounded  potential forbids the magnetic field to change the direction in the classical mechanics.  The field is reversed in the  quantum  mechanics.  The present tunneling  is distinct from the ones over finite potential barrieies   Ref.~\cite{Tunneling}.  Spontaneous magnetization emerges as an intrinsic property of the electron fluid, naturally linking itinerant ferromagnetism to the fundamental framework of gauge field theory.
This perspective unifies diverse manifestations of itinerant magnetism and suggests that the mechanism may operate in a wide range of metallic ferromagnets.

Recent studies have revealed intriguing connections between magnetism and the topology of electronic bands.
The so-called modern theory of orbital magnetization discussed in Ref.~\cite{ModernTheoryOfOrbitalMagnetism} has successfully provided an unambiguous definition of orbital magnetization within the framework of band theory; however, it deals with the induced magnetization obtained within the linear-response approximation, rather than spontaneous magnetization.

The concept of altermagnetism proposed by Ref.~\cite{Altermagnets} introduces a novel type of magnetic order and is of great interest, yet it relies essentially on specific band structures, such as those involving strong spin-orbit coupling.
In contrast, the mechanism proposed in the present work and in Ref.~\cite{Ishikawa-NPB} 
does not depend on the details of band structures, nor even on the existence of the spin degree of freedom. Universal $\log H$ term in the free energy has the origin in the spatial size  of the wave function that is inversly proportional to the magnetic field, and leads unusual adiabatic motions.   
It demonstrates that ferromagnetism can emerge as a manifestation of fundamental properties of gauge field theory, that is, the effective theory which distill the essence of electron fluid, thereby setting it apart from these earlier approaches by Refs.~\cite{ModernTheoryOfOrbitalMagnetism,Altermagnets}.

This work is partially supported by a Grant-in-Aid for Scientific Research (B) (Grant No. 24340043 and Grant No. 26K00623 ).  J. Goryo was collaborating in the eary stage of this research. The author  is grateful to J. Goryo,   H. Amitsuka,  and  H. Matsuyama for many useful discussions. 

 
\appendix
\section{Change of independent variables }
For slow motion of the magnetic field, two actions 
\begin{eqnarray}
& &S_1=\int_{t_1}^{t_2}dt [F(H(t)], \nonumber \\
& &S_2= \int_{t_1}^{t_2} dt[ F(H(t))+\frac{\partial }{\partial t} (AB )], 
\end{eqnarray}
where $A$ and $B$ are  functions or (functionals) of rellevant  variables are equivalent in the bulk region  $t_1 < t< t_2$. 
The last tem of $S_2$ is the difference between two actions and does not depends on the variable in the inside $t_1 < t< t_2$.  Because the differences are only in the boundaries, $S_1$ and  $S_2$ represent the same mechanical system.
We write the  variation of $S_1$ as 
\begin{eqnarray}
& &\delta S_1=\int_{t_1}^{t_2}dt  \frac{ \partial F(H(t)}{\partial H} \frac{\partial H}{\partial t} =\int_{t_1}^{t_2} \frac{ \partial F(H(t)}{\partial H} \delta { H},
\end{eqnarray}
and have  the equation of motion from $\frac{ \delta S_1}{\delta H}=0$  
\begin{eqnarray}
& & \frac{ \partial F(H(t))}{\partial H} =0. 
\end{eqnarray}
The variation of $S_2$ is also written as
\begin{eqnarray}
& &\delta S_2
=\int_{t_1}^{t_2}  [ \frac{\partial F}{\partial H}   \delta  H +A \delta B +  B \delta A ]. 
\end{eqnarray}
Now we choose 
\begin{eqnarray}
A=-H(t), B=\frac{\partial F}{\partial H}
\end{eqnarray}
then 
\begin{eqnarray}
\delta S_2=-\int_{t_1}^{t_2}  [ H   \delta B   ]. 
\end{eqnarray}
$S_2$ describes the evolution of $B$, and  is stational when $B$ satisfies the equation of motion, $\frac{ \delta S_2}{\delta B}=0$. This  is expressed by
\begin{eqnarray}
  H(B)  =0.  
\end{eqnarray}
$S_1(H)$ shows the motion of $H$ with the force  $ \frac{\partial F(H)}{ \partial H}$,  whereas $S_2(B)$ shows the motion of $B$ with the (effective) force  $H$. If the value of $H$ is measured directly, its motion is determined by  $S_1$. On the other hand, if the system is controlled by the force $B$,  its motion is determined by $S_2$.   
Now variation of 
\begin{eqnarray}
g(B)=BH-F(H)
\end{eqnarray}
is given by 
\begin{eqnarray}
\Delta  g(B)=H \Delta B +B \Delta H -\frac{\partial F}{\partial H} \Delta H=H \Delta B,
\end{eqnarray}
and is equivalent to $-\delta S_2$. $g(B)$ shows  the system's motion with  the change of $B$.

A system discribed by an external variable $O$ and free energy, 
\begin{eqnarray}
 f(O)&=&   -C_0 \log \frac{|O|}{O_0}+ \sum_l \frac{C_n}{n} {\bs O}^n, 
 \end{eqnarray}
behaves in the equivalent manner as Eq.$(\ref{free-energy}) $ at $O \sim 0$. The present results  provide  universal features  of the first order phase transition in systems of logarithmic term in the free energy. 
\section{Photon zero-point energy, and tunneling over the potential barrier  }
(1) Photon and phonon are bosons and described by symmetric many-body state.  Each mode has zero-point energy that is proportional to the frequency $\frac{1}{2} \hbar \omega$. The ground state energy   is expressed  
by these energies as         
\begin{eqnarray}
E_g=\frac{\hbar}{2}[ \int d {\bs k}   k_{photon}( \bs k)+ \int d {\bs q}   \omega_{photon}(\bs q)].
\end{eqnarray}
and contribute the free energy. The photon  gets a mass from  the Chern-Simons term and the zero point energy of  wave length longer than atomic distance   shifts.  The phonon spectrum depends on the  crystall's symmetry. The ground state  energy  of phonons depends on the lattice structure. These energies  are added to the the electron's energy 
studied in this paper.  
   
   (2)  
Free energy of the magnetic field Eq.$(\ref{free_B})$ depends on the integration path owing to the singulatity at $H=0$. In the present paper, we determines the free energy of the  process where the field is varied  adiabatically from   positive to negative values.  The wave functions  at $H>0$ are superpositions of left moving  and right moving waves, and at $H<0$ is   left moving  from the quantum tunneling.   The  system is described by 
\begin{eqnarray}
H_{eff}=\frac{1}{2m}P_H^2-f(H).
\end{eqnarray}
Here $m$ and $P_H$ are the effective mass and momentum conjugate to $H$  derived from  the effective Lagrangian.    The  transmitted   energy  over the  barrier, $\Delta$, is  expressed   from  a semiclassical approximation as, 
\begin{eqnarray}
\Delta=\frac{1}{2} \sqrt{\frac{1+\kappa }{m}}e^{-   \sqrt{2 \pi m } C_0 \frac{4 \pi\hbar}{e}      } .
\end{eqnarray}
This gives the difference of energy  
\begin{eqnarray}
  C_0 \log\frac{H_0({+})}{H_0({-})} =2 \Delta
\end{eqnarray}
   in regions   $H>0$ and  $H<0$, and it follows that  $f_{+}(|H|) = f_{-}(|H|) +2 \Delta$.  The tunneling does not take place in going down  the energy hill. The enegy is asymmetric and breaks  the time reversal invariance.

 The tunneling energy may have corrections if  the  energy from the long-range fluctuation is included. This energy was found to exist,    \cite{Ishikawa-Tobita_{PTEP},Ishikawa-Takesada}, and   makes  
   the energy of the final state $E_f=E_i-\Delta E$,  where $\Delta E \neq 0$.  $\Delta E >0$ in many situations. Major effects of the present quantum phase transition are not modified.



\begin{thebibliography}{99}

\bibitem{Wolf} S. A. Wolf \textit{et al.}, Science \textbf{294}, 1488 (2001).

\bibitem{Zutic} \v{Z}uti\'c, J. Fabian, and S. Das Sarma, Rev. Mod. Phys. \textbf{76}, 323 (2004).


\bibitem{Stoner} E. C. Stoner, Proc. R. Soc. A \textbf{165}, 372 (1938).

\bibitem{Hubbard} J. Hubbard, Proc. R. Soc. A \textbf{276}, 238 (1963).

\bibitem{Moriya} T. Moriya, Spin Fluctuations in Itinerant Electron Magnetism (Springer, 1985).

\bibitem{Georges} A. Georges \textit{et al.}, Rev. Mod. Phys. \textbf{68}, 13 (1996).

\bibitem{Held} K. Held, Adv. Phys. \textbf{56}, 829 (2007).

\bibitem{Stoner_2} E. C. Stoner and E. P. Wohlfarth.Philos.  R. Soc. London Ser.A \textbf{240}, 399 (1948)
\bibitem{Ishikawa-NPB} K. Ishikawa, Nucl. Phys. B {\bf 1007}, 116663 (2024). 
The CS term in the closed system  was provided. Note that this agrees with the Hall conductivity $\sigma_{xy}$ in the gapped system.  \cite{ Ishikawa-Matsuyama, Imai-Ishikawa-Matsuyama-Tanaka} 
  

\bibitem{Ishikawa-Matsuyama} K. Ishikawa and T. Matsuyama, Z. Phys. C \textbf{33}, 41 (1986).

\bibitem{Imai-Ishikawa-Matsuyama-Tanaka} N. Imai, K. Ishikawa, T. Matsuyama, and I. Tanaka, 
Phys. Rev. B \textbf{42}, 10610 (1990).

\bibitem{Footnote-charge} In this unit, the electric charge $e=\sqrt{4 \pi \alpha}$. 

\bibitem{Orthogonality} K. Ishikawa, Prog. Theor. Phys. \textbf{55}, 588 (1976).

\bibitem{Mahan} G. D. Mahan, {\it Many-Particle Physics}, 3rd ed. (Springer, New York, 2000).
\bibitem{Sin-Itiro-Tomonaga} Sin-Itiro Tomonaga, 
Nobel Lecture: Development of Quantum Electrodynamics, Nobel Foundation (1965).


\bibitem{Thouless-Valatin} D. J. Thouless and J. G. Valatin, Nuclear Physics \textbf{31}, 211 (1962). 




\bibitem{von-Neumann} K. Ishikawa, N. Maeda, and H. Suzuki, Physica E \textbf{4}, 37 (1999).

\bibitem{Footnote-symmetric-gauge}
Using the symmetric gauge $\vec{A}=(- y B/2, x B/2)$, 
we obtained the equivalent result without regularization, since the translational symmetry is broken explicitly.  
Although it may seem peculiar that the need for regularization depends on the gauge choice, the result is justified 
because it agrees with the gauge-invariant von Neumann lattice field theory~\cite{Ishikawa-NPB, von-Neumann}, which 
needs no regularization. This also underscores the essential role of the field-theoretic description.



\bibitem{Footnote-variation-free-energy} The variation of the free energy functional 
means $\beta^{-1}$ times the effective action for the external gauge field (gauge-flux insertion) 
$\delta A_\mu$, which causes the adiabatic change of chemical potential $\mu$ and magnetization $B$.  

\bibitem{Footnote-TF} In a gapless electron system, the effective Lagrangian exhibits Thomas--Fermi-type screening. Although this may superficially resemble a Proca mass and seem to imply a breaking of gauge invariance, the Ward identity ensures that gauge invariance is fully preserved. The term simply reflects the breaking of Lorentz invariance by the medium's preferred rest frame, rather than any violation of gauge symmetry. 
Thus, Thomas--Fermi screening should be viewed as the suppression of electrostatic interactions in a medium, not as a symmetry-breaking effect.

\bibitem{Footnote-B-linear} 
Linear term in $H$ may appear, but is ginored here for simplicity.
\bibitem{Schwinger} J. Schwinger,Phys. Rev, \text{73} 416-417, (1948), 
\bibitem{Footnote-coercive}. $H_c$ becomes $10^5 \text A/m$ if $\kappa=10^5$ and $\mu n_e=0.1 eV \times 8.5 \times 10^{28}  m^{-3}$ are substituted.  
\bibitem{Ishikawa-Tobita} K. Ishikawa and Y.Tobita, Prog. Theor. Phys. \textbf{122}, 1111 (2009).

\bibitem{Ishikawa-Nishio} K.Ishikawa and Y.Nishio, Annals. of Physics, 469,2024,169750

\bibitem{Belitz-PRL99} D. Belitz, T. R. Kirkpatrick, and T. Vojta,
Phys. Rev. Lett. 82, 4707 (1999).

\bibitem{Brando-RMP} M. Brando, D. Belitz, F. M. Grosche, and T. R. Kirkpatrick, Rev. Mod. Phys. 88, 025006 (2016).

\bibitem{Goko-npjQM} T. Goko et al., npj Quantum Mater. 2, 44 (2017).

\bibitem{Kittel} C. Kittel, Rev. Mod. Phys. \textbf{21}, 541 (1949).

\bibitem{Footnote-wall-width} Typically, $l_\parallel \simeq l_{TF}=\varepsilon^{1/2} \kappa_{TF}^{-1/2}\simeq 1\AA$.  

\bibitem{Fe3GeTe2} C. Tan {\it et al.}, Nat Commun. \textbf{9}, 1554 (2018). 

\bibitem{Carol-Field-Jackiw} S. M. Carroll, G. B. Field and R. Jackiw, Phys. Rev. D \textbf{41}, 1231 (1990).

\bibitem{Tunneling}  
A.O.Caldelra and A.J. Leggett, Phys. Rev. Lett.  \textbf{46}, 211 (1981). 

\bibitem{ModernTheoryOfOrbitalMagnetism} See, for a review, 
R. Resta, J. Phys.: Condens. Matter \textbf{22}, 123201 (2010). 

\bibitem{Altermagnets} L. \v{S}mejkal, J. Sinova, T. Jungwirth, \textit{et al.}, 
Phys. Rev. X \textbf{12}, 040501 (2022).
\bibitem{Ishikawa-Tobita_{PTEP}} K. Ishikawa and Y.Tobita, Prog. Theor. Exp. Phys. \textbf{2013},073B02,https://doi.org/
\bibitem{Ishikawa-Takesada} K. Ishikawa and M.Takesada, ``New class of quantum transitions exbiting large-scale intercorrelations''in preparation. 
\end{thebibliography}
\end{document}